\documentclass[a4paper,11pt]{article}
\usepackage{pos}
\usepackage{bm}
\usepackage{natbib}
\title{Modelling the Radio Emission of Inclined Air Showers in the 50-200 MHz Frequency Band for GRAND}
\ShortTitle{Modelling the Radio Emission of Inclined Air Showers in 50-200 MHz}

\author*[a]{Lukas Gülzow}
\author[a,b]{Tim Huege}
\author[c,d]{Kumiko Kotera}
\author[e]{Olivier Martineau}
\author[a]{Markus Roth}
\author[f]{Felix Schlüter}

\affiliation[a]{Institute for Astroparticle Physics (IAP), Karlsruhe Institute of Technology, Karlsruhe, Germany}
\affiliation[b]{ Astrophysical Institute, Vrije Universiteit Brussel, Brussels, Belgium}
\affiliation[c]{Sorbonne Université, CNRS, Institut d’Astrophysique de Paris, Paris, France}
\affiliation[d]{Vrije Universiteit Brussel (VUB), Dienst ELEM, Pleinlaan 2, B-1050, Brussels, Belgium}
\affiliation[e]{Sorbonne Université, CNRS, Laboratoire de Physique Nucléaire et de Hautes Energies (LPNHE), Paris, France}
\affiliation[f]{Université Libre de Bruxelles, Science Faculty CP230, B-1050 Brussels, Belgium}

\emailAdd{lukas.guelzow@kit.edu}

\abstract{The Giant Radio Array for Neutrino Detection (GRAND) is a distributed, sparse ground antenna array designed to detect the radio emission from highly inclined extensive air showers induced by ultra-high-energy particles in the atmosphere.
We use CoREAS air-shower simulations to adapt an existing signal model of the radio emission of inclined showers to the $50-200\,\mathrm{MHz}$ frequency band GRAND is sensitive to.
The model uses a parameterisation of the charge excess emission to isolate the geomagnetic component.
By fitting a one-dimensional lateral distribution function to the geomagnetic energy fluence of a shower, we reconstruct its radiation energy.
This work details the signal model and the intrinsic event reconstruction of our method, as well as the adaptations to the new frequency band. This work is part of the NUTRIG project.}

\FullConference{%
  10th International Workshop on Acoustic and Radio EeV Neutrino Detection Activities - ARENA2024\\
  June 10-14 2024\\
  University of Chicago, Chicago, USA
}

\begin{document}
\maketitle

\section{Introduction}
\label{sec:intro}
This project builds upon the established signal model and energy reconstruction of inclined air showers developed for the AugerPrime Radio detector in the frequency band of $30-80\,\mathrm{MHz}$~\cite{Model-paper}. 
The radio emission of air showers is composed of both geomagnetic and charge excess components~\cite{Huege-review}.
For inclined air showers, the emission pattern exhibits geometry-dependent asymmetries.
The model corrects for these asymmetries with a so-called early-late correction~\cite{Huege:2018kvt}.
The fraction of the charge excess $a_\text{ce}$ of the total emission is described with a parameterisation and used to isolate the geomagnetic component of the energy fluence $f_\text{geo}$.
A symmetric geomagnetic lateral distribution function (LDF), composed of the sum of a Gaussian and a sigmoid, is fitted to $f_\text{geo}$.
By integrating the fitted LDF, the geomagnetic radiation energy $E_\text{geo}$ of the shower is calculated.
This allows a reconstruction of the electromagnetic energy $E_\text{em}$.

To enable the energy reconstruction for sparse antenna arrays, the shape parameters of the LDF are described using their behaviour with respect to the distance $d_\text{max}$ to the shower maximum.
This method reduces the amount of free parameters to only $d_\text{max}$, the geomagnetic radiation energy $E_\text{geo}$ and the two core coordinates.
On simulations with a sparse antenna grid, this energy reconstruction achieves an intrinsic $<5\%$ resolution with no significant biases~\cite{Model-paper}. For fully realistic simulations, exclusively using information available in real measurements, the resolution only decreases to $\sim 6\%$ while remaining free of biases~\cite{Huege-performance}.

In this work, we adapt the above signal model to the $50-200\,\mathrm{MHz}$ frequency band in which the Giant Radio Array for Neutrino Detection (GRAND) operates \cite{GRAND:2018iaj}.
To tune the model, we use simulation libraries generated with CoREAS~\cite{Huege:2013vt} with star-shaped antenna grids for highly inclined air showers at the GRAND@Auger site in Argentina and the GP300 site in China \cite{Chiche_arena_pos, Kotera_arena_pos}.
The library for the site in Argentina is the same as used in reference~\cite{Model-paper}, filtered to $50-200\,\mathrm{MHz}$.
Both libraries contain $\sim 4\,000$ air shower simulations with 240 stations each and have zenith angles ranging from 65 to 85 degrees.

In Section~\ref{sec:changes}, we cover the changes made to the signal model for the $50-200\,\mathrm{MHz}$ frequency band and what warrants them.
In Section~\ref{sec:comparison}, we go into a detailed comparison of the features of the signal model and the intrinsic energy reconstruction performance of our method between GRAND’s two sites in Argentina and China.
Finally, we draw our conclusions in Section~\ref{sec:conclusions}.

\section{Changes to the signal model}
\label{sec:changes}

\subsection{Fit of the charge excess fraction of the total energy fluence}
\label{sec:ce-fraction}
We isolate the geomagnetic energy fluence $f_\text{geo}$ by using the $\Vec{v}\times\Vec{B}$ polarisation of the electric field and a parameterisation of the charge excess fraction of the total emission.
As a result, we find the symmetric fluence pattern of the geomagnetic emission.
The parameterised geomagnetic fluence $f_\text{geo}^\text{par}$ and the parameterisation of the charge excess fraction $a_\text{ce}$ are given by eqs. (4.10) and (4.12) in reference \cite{Model-paper}.

\begin{figure}
\begin{center}
	\includegraphics[clip, trim=13.5cm 0.5cm 0.5cm 3.5cm, width=0.85\columnwidth]{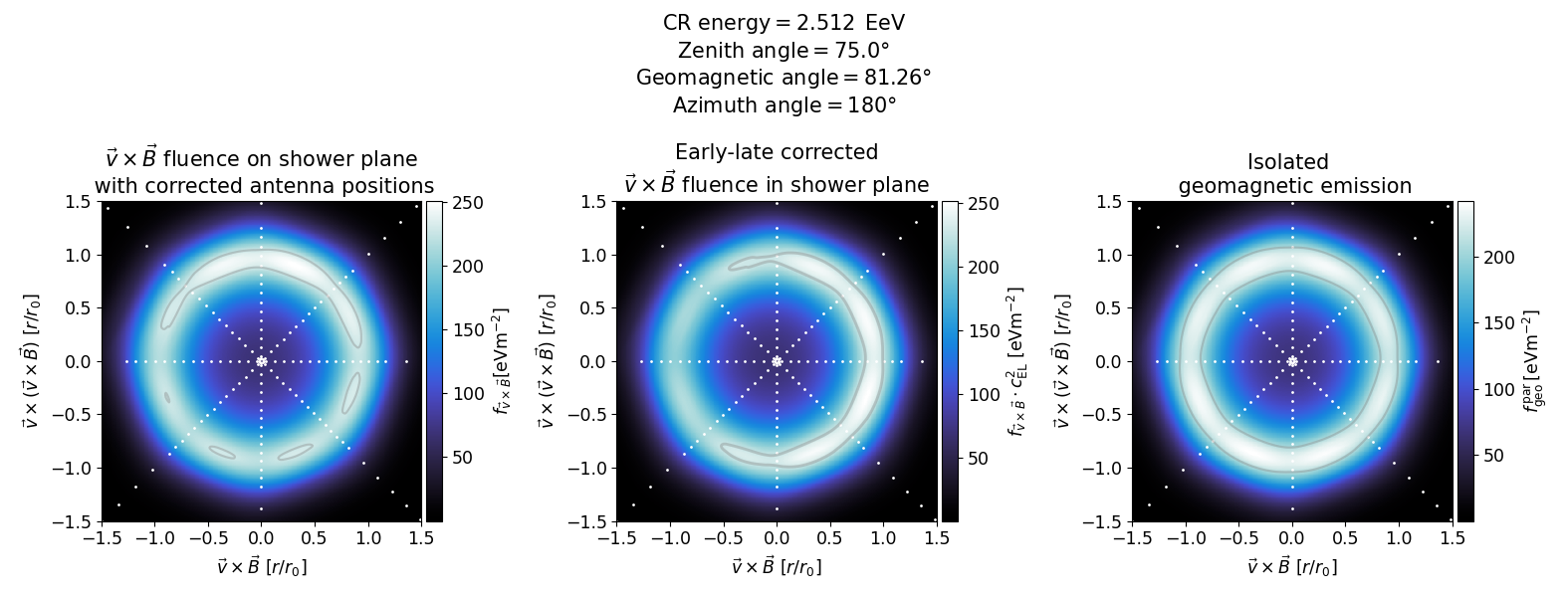}
    \caption{The first two steps of our signal model which eliminate geometric asymmetries and to isolate the geomagnetic energy fluence, respectively. Shown for an example simulation of the GRAND@Auger site with a star-shaped antenna pattern (white dots). The regions with $90\%$ or more of the maximum energy fluence are outlined in grey. Axis distances $r$ are normalised with respect to the Cherenkov radius $r_0$. \textbf{Left:} Early-late correction factors $c_\text{EL}$ applied to energy fluence and antenna positions. \textbf{Right:} Parametric geomagnetic fluence $f_\text{geo}^\text{par}$.}
    \label{fig:footprints}     
\end{center}
\end{figure}

In Fig.~\ref{fig:footprints}, we show the steps of the signal model up to this point for the $50-200\,\mathrm{MHz}$ frequency band on the emission pattern in the shower plane.
We show the early-late corrected $\Vec{v}\times\Vec{B}$ energy fluence in the left panel.
In the right panel, the now symmetric geomagnetic fluence is displayed.
\subsection{Lateral distribution function of the geomagnetic energy fluence}
\label{sec:LDF}
The LDF $f_\text{GS}(r)$ of the geomagnetic fluence from the signal model described in reference \cite{Model-paper},
\begin{equation}
f_\text{GS}(r)=\frac{E_\text{geo}}{E_0} \left[\text{exp}\left(-\left(\frac{r-r_0}{\sigma}\right)^{p(r)}\right)+\frac{a_\text{rel}}{1+\text{exp}\left(s\cdot\left(\frac{r}{r_0}-r_{02}\right)\right)}\right],
\end{equation}
is composed of a Gaussian and a sigmoid term.
It has six shape parameters: the Gaussian peak position $r_0$, the Gaussian width $\sigma$, the Gaussian exponent $p(r)$, the relative amplitude of the sigmoid compared to the Gaussian $a_\text{rel}$, the sigmoid slope $s$, and the sigmoid length scale $r_{02}$.
Additional free parameters are the geomagnetic radiation energy $E_\text{geo}$ (with normalisation $E_0$) and the position of the symmetry center of the radio-emission footprint ("radio core").

We make changes to the LDF functional form to adapt it to the $50-200\,\mathrm{MHz}$ frequency band.
Due to the wider frequency band, the total intensity of the radio emission increases.
This added intensity is mainly concentrated on the Cherenkov ring. As such, the amplitude on the ring is much larger compared to its inside.
To account for this change, we introduce a new parameter $p_\text{inner}$ which allows the slope of the ``Gaussian'' to become steeper for $r<r_0$, 
\begin{equation}
f_\text{LDF}(r)=
    \begin{cases}
        \frac{E_\text{geo}}{E_0} \left[\text{exp}\left(-\left(\frac{|r-r_0|}{\sigma}\right)^{p_\text{inner}}\right)+\frac{a_\text{rel}}{1+\,\text{exp}\left(s\cdot\left(\frac{r}{r_0}-r_{02}\right)\right)}\right] & \text{for} \quad r<r_0,\\[15pt]
        \frac{E_\text{geo}}{E_0} \left[\text{exp}\left(-\left(\frac{r-r_0}{\sigma}\right)^{p(r)}\right)+\frac{a_\text{rel}}{1+\,\text{exp}\left(s\cdot\left(\frac{r}{r_0}-r_{02}\right)\right)}\right] & \text{for} \quad r\geq r_0.
    \end{cases} 
\label{eq:new_LDF}
\end{equation}
For $r<r_0$, we fit the exponent $p_\text{inner}$ directly as a scalar constant. For $r\geq r_0$, $p(r)$ remains as described in reference~\cite{Model-paper}: $p(r)$ is allowed to decrease from 2 to better describe the tail of the LDF.

\begin{figure}
\begin{center}
	\includegraphics[clip, trim=2cm 0cm 0cm 1.34cm, width=0.85\columnwidth]{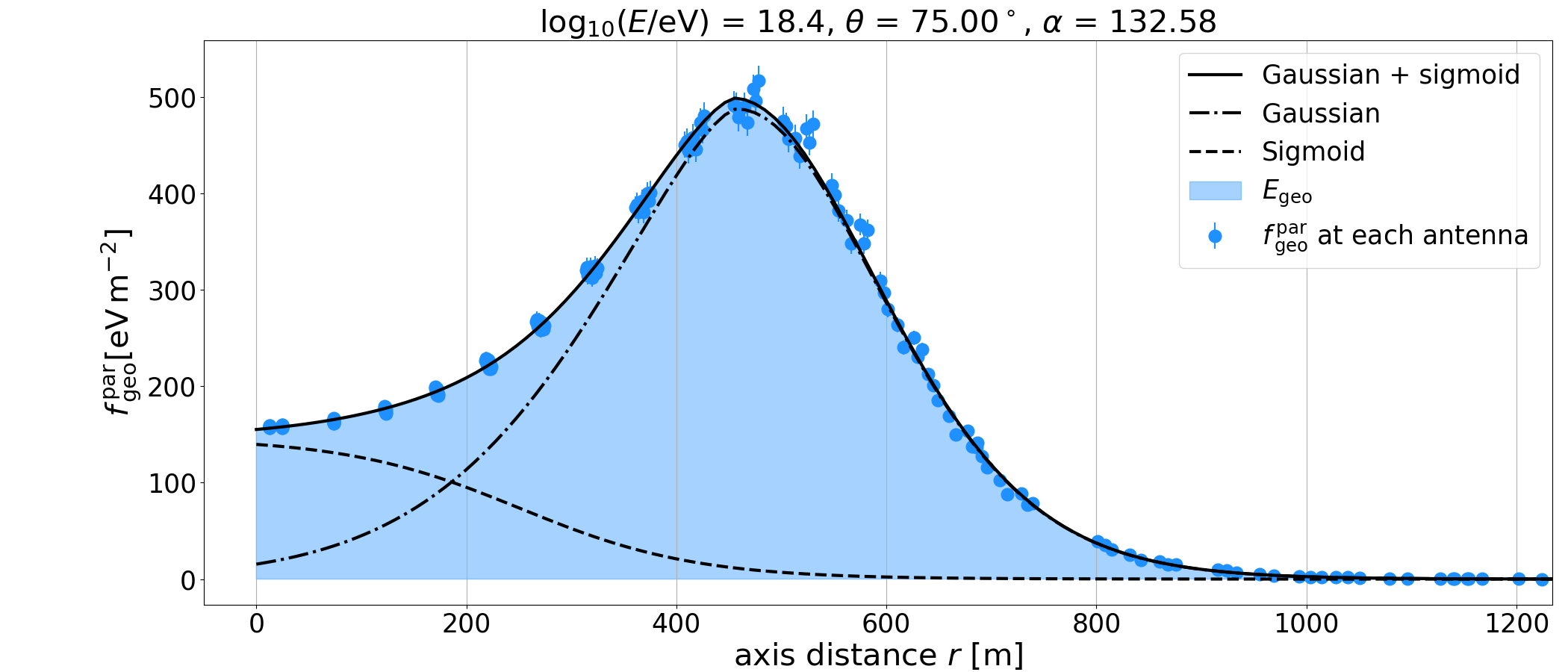}
    \caption{Fit of the new LDF $f_\text{LDF}(r)$ to the geomagnetic fluence $f_\text{geo}^\text{par}$ of an example simulation of the GP300 site with a star-shaped antenna pattern. $f_\text{geo}^\text{par}$ is shown as blue points and the LDF as a solid black line. The coloured area under the curve represents the geomagnetic radiation energy $E_\text{geo}$.}
    \label{fig:integral}
\end{center}
\end{figure}

The new LDF has seven free shape parameters in addition to $E_\text{geo}$ and the radio core position.
Fig.~\ref{fig:integral} shows an example of the LDF fit to $f_\text{geo}^\text{par}$ from a simulation in the $50-200\,\mathrm{MHz}$ frequency band. 
The area under the curve represents the geomagnetic radiation energy $E_\text{geo}$.

The parameter $r_0$ closely matches the Cherenkov radius.
In the lower-frequency model~\cite{Model-paper}, $r_0$ is fixed directly as the radius of the Cherenkov cone under the assumption that the emission originates from the shower maximum $X_\text{max}$.
The function for $r_0$ uses the atmospheric model and the opening angle of the Cherenkov cone at the position of $X_\text{max}$, corresponding to $d_\text{max}$, to predict the Cherenkov radius $r_0^\text{pred}$.
Any deviations are sufficiently compensated for by the other parameters~\cite{Model-paper}.

For higher frequencies, we find a consistent deviation of a free fit of $r_0$ from the predicted Cherenkov radius $r_0^\text{pred}$ with respect to $d_\text{max}$.
We accommodate this feature by parameterising $r_0$ with $d_\text{max}$ as
\begin{equation}
r_0=r_0^\text{pred}\cdot \left(c_1 + c_2\cdot d_\text{max} + \frac{c_3}{d_\text{max}^2}\right).
\end{equation}

In addition, we make an optimisation to the fit of the radio core position.
The radio core is shifted from the shower core for inclined showers due to refractive displacement~\cite{Schluter-refr_displacement}.
A free fit struggles to reliably converge on the true core for the higher frequency band. 
We constrain the core fit by fixing the possible core positions to a line defined by the projection of the shower axis on the ground plane.
This prevents shifts to the side which do not have a physical basis.

\section{Comparison between sites in Argentina and China}
\label{sec:comparison}

\subsection{New fits of the charge excess fraction}
\begin{figure}
	\includegraphics[clip, trim=13cm 0.8cm 0cm 0.7cm, width=0.5\columnwidth]{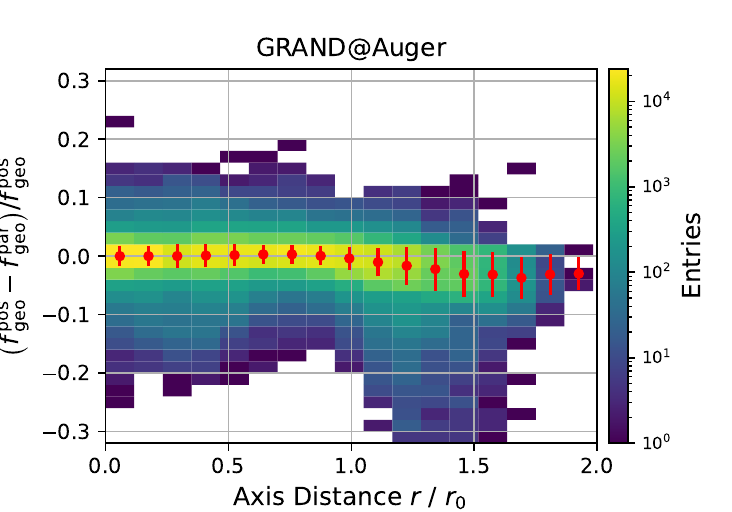}
    \includegraphics[clip, trim=12.5cm 0.8cm 0.5cm 0.7cm, width=0.5\columnwidth]{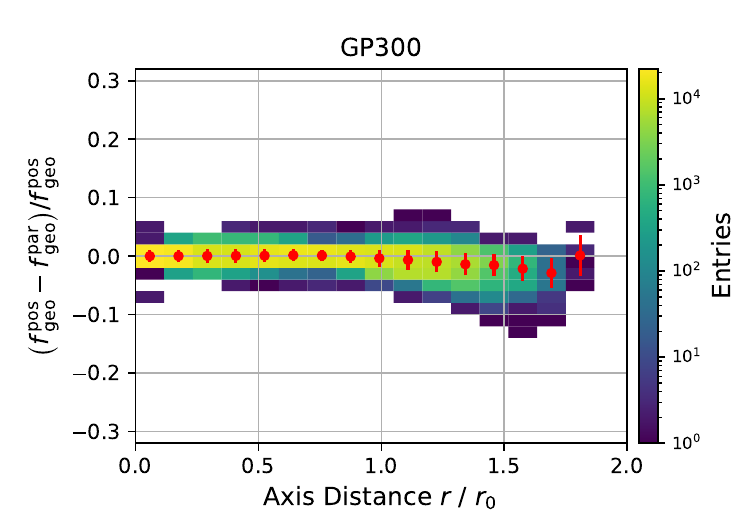}
    \caption{2D Histogram of the deviations between $f_\text{geo}^\text{par}$ and $f_\text{geo}^\text{pos}$ against the station-to-axis distance in units of Cherenkov radius $r_0$ for individual stations. The colour map shows the number of stations in each bin. Red points and error bars represent the mean and standard deviation in each column. Stations affected by thinning have been removed. \textbf{Left:} GRAND@Auger site in Argentina. \textbf{Right:} GP300 site in China.}
    \label{fig:ce_fraction_comp}
\end{figure}
We refit the charge excess fraction for both GRAND sites for the new frequency band. 
For GRAND@Auger, the site characteristics stay the same as in reference \cite{Model-paper}.
For the GP300 site in China, we use a different atmospheric model, observation level, and geomagnetic field \cite{GRAND:2018iaj}.
The charge excess emission in simulations only varies slightly depending on atmospheric model.
However, a different geomagnetic field leads to significant changes between charge excess fractions for the Auger and GP300 sites, respectively,
\begin{equation}
a_\text{ce}^\text{ARG}=\left[0.302-\frac{d_\text{max}}{729\,\mathrm{km}}\right]\cdot\frac{r_\text{axis}}{d_\text{max}}\cdot\text{exp}\left(\frac{r_\text{axis}}{682\,\mathrm{m}}\right)\cdot\left[\left(\frac{\rho_\text{max}}{0.422\,\mathrm{kg}\,\mathrm{m}^{-3}}\right)^{2.98}+0.178\right],
\label{eq:a_ce_param_auger}
\end{equation}
\begin{equation}
a_\text{ce}^\text{CHN}=\left[0.229-\frac{d_\text{max}}{1106\,\mathrm{km}}\right]\cdot\frac{r_\text{axis}}{d_\text{max}}\cdot\text{exp}\left(\frac{r_\text{axis}}{614\,\mathrm{m}}\right)\cdot\left[\left(\frac{\rho_\text{max}}{0.668\,\mathrm{kg}\,\mathrm{m}^{-3}}\right)^{1.43}+0.166\right],
\label{eq:a_ce_param_china}
\end{equation}
since it affects the strength of the geomagnetic emission.

In Fig.~\ref{fig:ce_fraction_comp}, we evaluate the parameterised geomagnetic fluence $f_\text{geo}^\text{par}$ (see Section~\ref{sec:ce-fraction}) for both sites.
We compare $f_\text{geo}^\text{par}$ with $f_\text{geo}^\text{pos}$, the geomagnetic fluence determined with a position- and polarisation-based calculation from the electric field components in the $\Vec{v}\times\Vec{B}$ and $\Vec{v}\times\Vec{v}\times\Vec{B}$ polarisations (see eq. (3.2) in reference \cite{Model-paper}). 
Both methods to determine the geomagnetic fluence agree well with each other for both sites.
For axis distances larger than the Cherenkov radius $r_0$, we see less accuracy due to the charge excess fraction becoming very small.
The strong magnetic field in China also causes the charge excess fraction to decrease and leads to a loss of coherence at the highest zenith angles~\cite{Chiche:2024yos}.
Stations affected by thinning have been removed~\cite{Model-paper}.

\subsection{Determination of \texorpdfstring{$\bm{r_0}$}{r_0}}
\label{sec:r_0}
We show how $r_0$ is determined for the two sites, as outlined in Section~\ref{sec:LDF}. 
\begin{figure}
	\includegraphics[clip, trim=0.5cm 0.5cm 0cm 0.2cm, width=0.5\columnwidth]{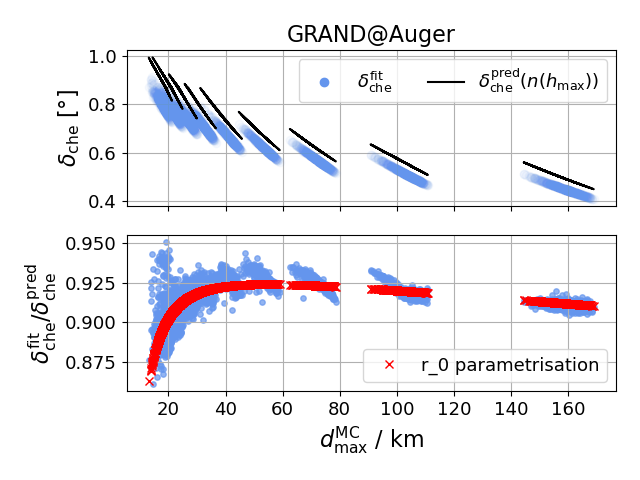}
    \includegraphics[clip, trim=0.5cm 0.5cm 0cm 0.2cm, width=0.5\columnwidth]{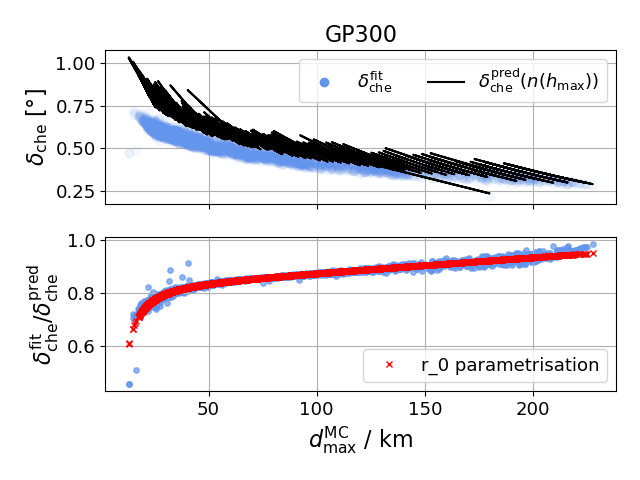}
    \caption{The left panel shows the GRAND@Auger site, the right panel the GP300 site. \textbf{Top row:} Comparison of the predicted opening angle of the Cherenkov cone $\delta^\text{pred}_\text{che}(n(h_\text{max}))$ (black lines) and the values determined in the LDF fit $\delta^\text{fit}_\text{che}$ (blue points). $n(h_\text{max})$ is the refractive index of the atmosphere at the height $h_\text{max}$ of the shower maximum. \textbf{Bottom row:} Deviation of $\delta^\text{fit}_\text{che}$ from $\delta^\text{pred}_\text{che}$ as blue points. The parameterisation for $r_0$ (see eqs.(\ref{eq:r_0_ARG}) and (~\ref{eq:r_0_CHN})) is fitted to the deviation as red crosses. }
    \label{fig:r_0}
\end{figure}
Fig.~\ref{fig:r_0} displays the prediction of the Cherenkov angle $\delta_\text{che}$ and the deviation of the fitted $\delta_\text{che}^\text{fit}$ against $d_\text{max}$. 
The parameterisations for $r_0$ are given by
\begin{equation}
r_0^\text{ARG}=r_0^\text{pred}\cdot \left(0.940 - \frac{d_\text{max}}{5931\,\mathrm{km}} + \frac{1361\,\mathrm{km}^2}{d_\text{max}^2}\right),
\label{eq:r_0_ARG}
\end{equation}
\begin{equation}
r_0^\text{CHN}=r_0^\text{pred}\cdot \left(0.823 + \frac{d_\text{max}}{1791\,\mathrm{km}} - \frac{4266\,\mathrm{km}^2}{d_\text{max}^2}\right),
\label{eq:r_0_CHN}
\end{equation}
for the GRAND@Auger and GP300 sites, respectively.

\subsection{Coherence loss, geomagnetic angle and density correction}
For the $50-200\,\mathrm{MHz}$ frequency band, we expect less coherence from
the radio emission which leads to lower signal fluences.
The site of GP300 in China also features a stronger magnetic field than at GRAND@Auger.
This contributes to the coherence loss, especially at the highest zenith angles, and will influence the polarisation signatures of the emission as well~\cite{Chiche:2024yos}.

In our model, this plays into the correction we apply to the radiation energy $E_\text{geo}$ to calculate the corrected radiation energy $S_\text{geo}$, which is independent of shower geometry.
The coherence loss leads to a new dependence on the geomagnetic angle $\alpha$, and causes $E_\text{geo}$ to fall off at low densities at the shower maximum $\rho_\text{max}$.
To account for this, we modify eq. (5.2) from reference \cite{Model-paper} with new parameters $c_\alpha$, $p_2$ and $p_3$.
This yields the following normalisation:
\begin{equation}
S_\text{geo}=\frac{E_\text{geo}}{\text{sin}^{c_\alpha}(\alpha)}\cdot\frac{1}{\left(1-p_0+p_0\cdot\text{exp}\left(p_1\cdot\left[\rho_\text{max}-\langle\rho\rangle\right]\right) - p_2\cdot \rho_\text{max}^{-1} + p_3\right)^2}.
\label{eq:density_corr}
\end{equation}

\begin{figure}
	\includegraphics[clip, trim=1.5cm 1cm 2cm 0.75cm, width=0.5\columnwidth]{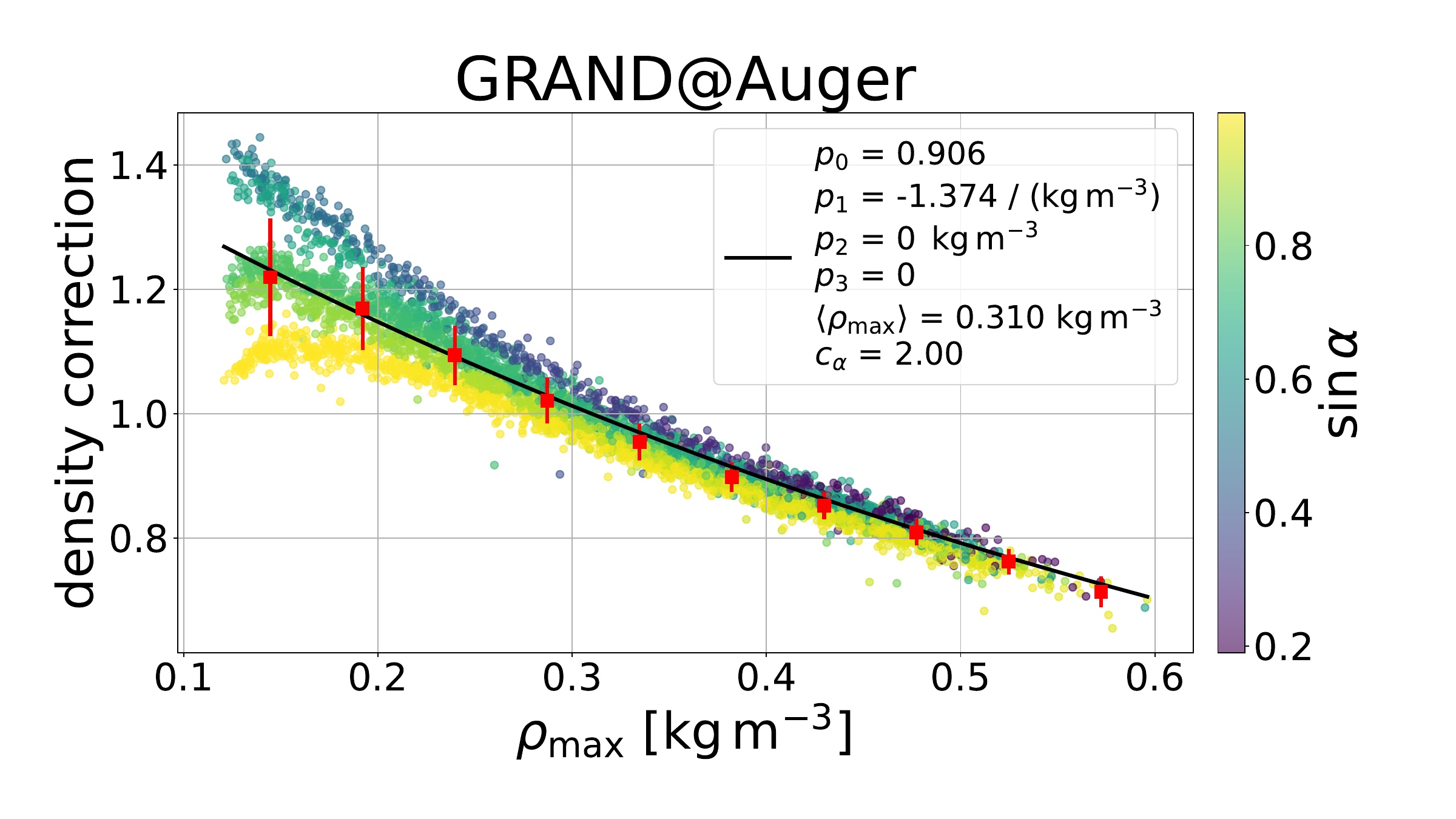}
    \includegraphics[clip, trim=1.5cm 1cm 2.4cm 0.75cm, width=0.5\columnwidth]{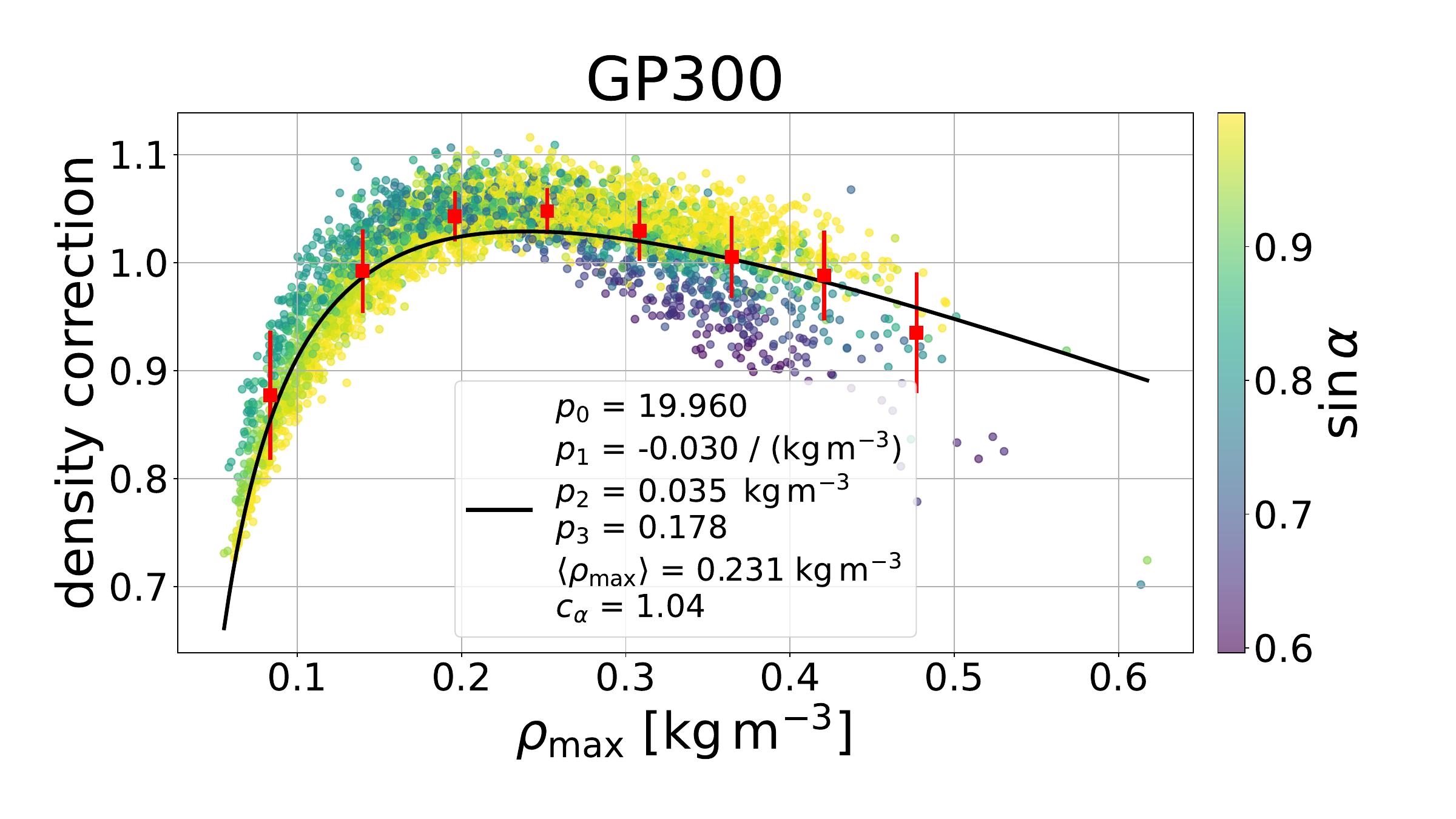}
    \caption{Modelled density correction for the calculation of $S_\text{geo}$ plotted against $\rho_\text{max}$. The colour map shows the dependency on the geomagnetic angle $\alpha$. The black line is fitted to the points to provide the function of the density correction for the energy reconstruction. Red points and error bars show the means and standard deviations, respectively. \textbf{Left:} GRAND@Auger. \textbf{Right:} GP300.}
    \label{fig:density_corr}
\end{figure}

Fig.~\ref{fig:density_corr} shows how the density correction behaves for the two sites.
For GRAND@Auger, we only observe slight effects of coherence loss for large values of $\text{sin}(\alpha)$ at low $\rho_\text{max}$ (i.e.~for high zenith angles).
As such, the new parameters do not have an influence.
However, the coherence loss is much more significant for the stronger magnetic field at the GP300 site.
\subsection{Energy reconstruction performance}
\label{sec:recon}
From the corrected radiation energy $S_\text{geo}$, we calculate the electromagnetic energy $E_\text{em}$ of the shower using a power law with spectral index $\gamma$ as
\begin{equation}
E_\text{em}=10\,\mathrm{EeV}\left(\frac{S_\text{geo}}{S_{19}}\right)^\frac{1}{\gamma}.
\label{eq:em}
\end{equation}
Here, $S_{19}$ can be understood as the geomagnetic radiation energy of a $10\,\mathrm{EeV}$ ($=10^{19}\,\mathrm{eV}$) air shower.

\begin{figure}
	\includegraphics[width=0.5\columnwidth]{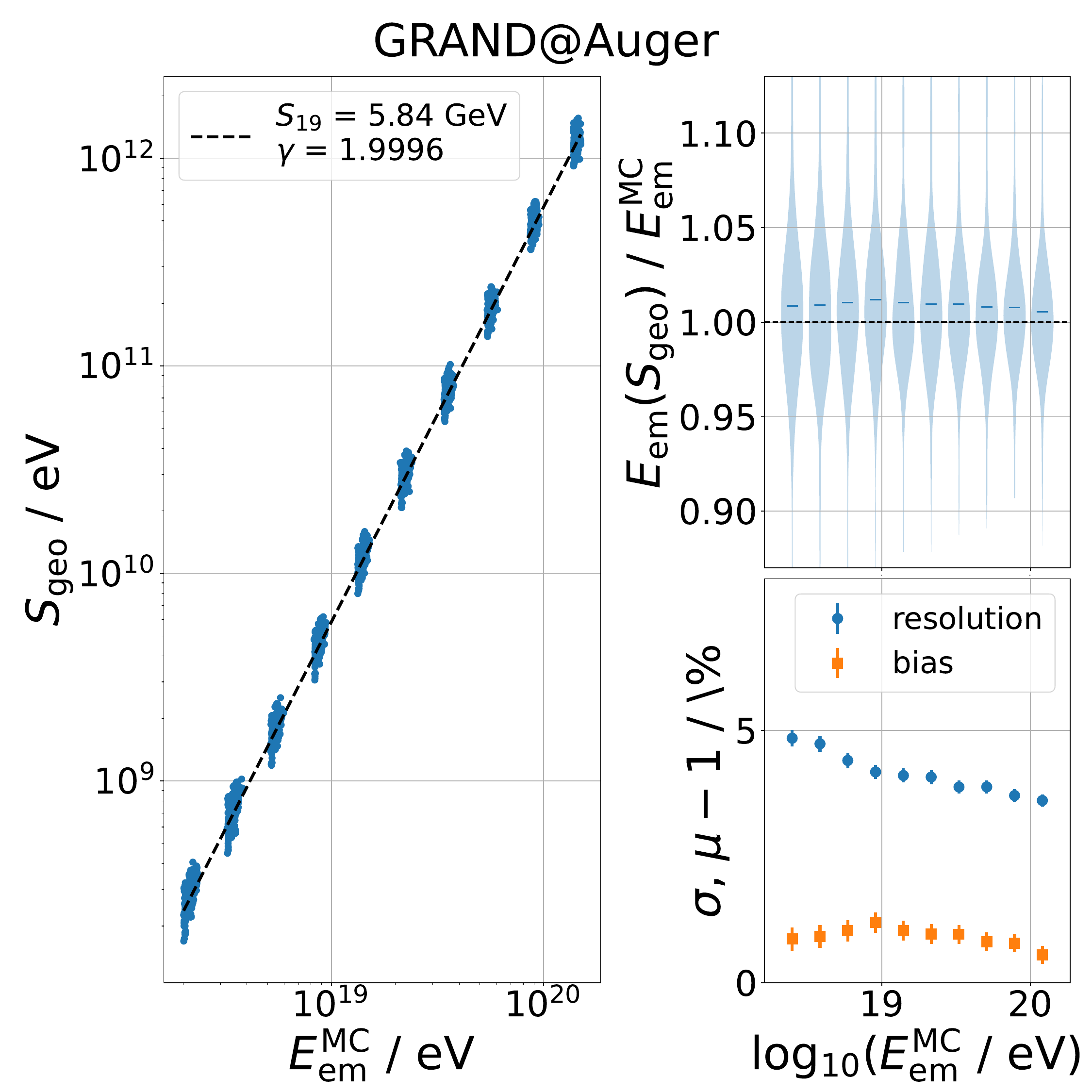}
     \includegraphics[width=0.5\columnwidth]{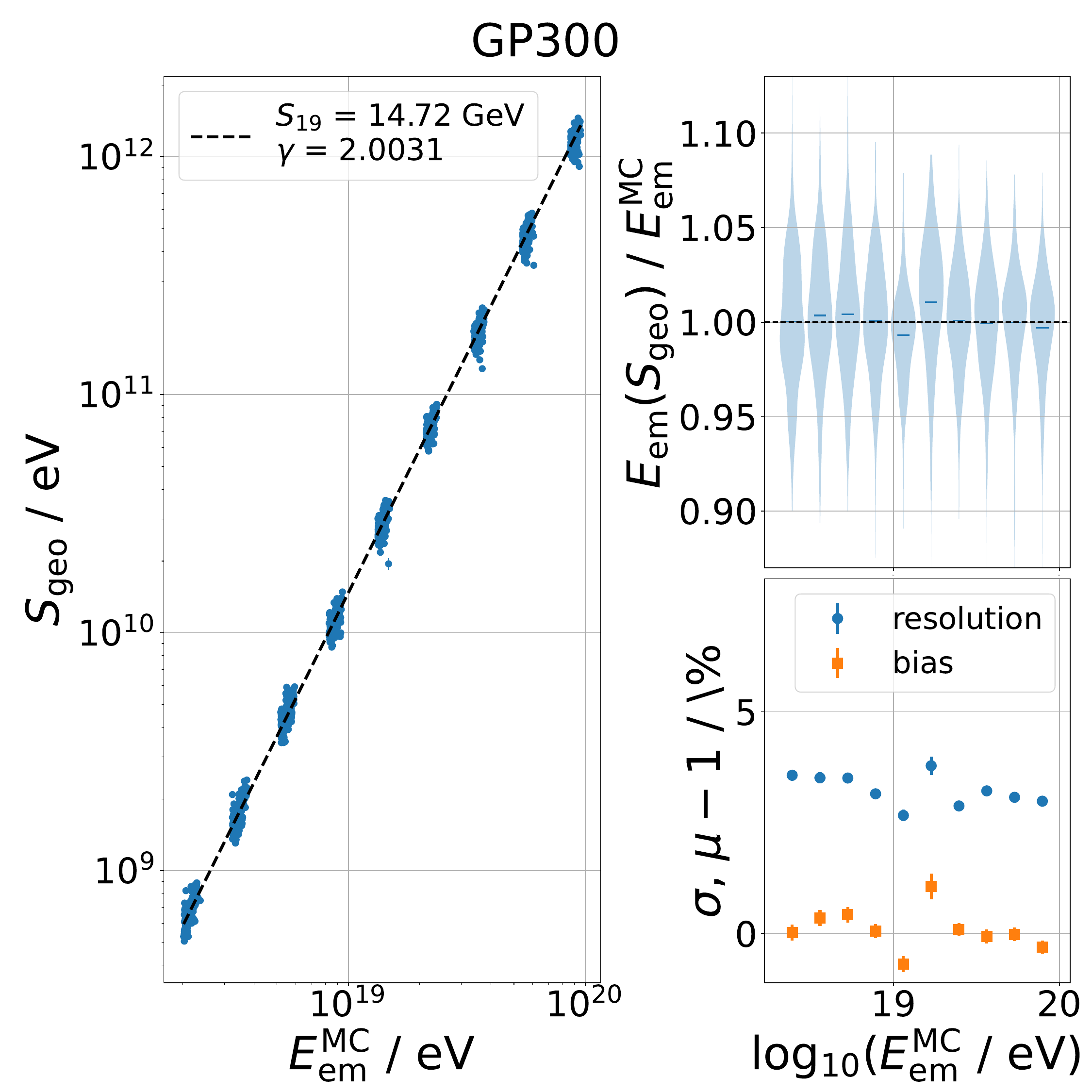}
    \caption{The left column of each panel shows the reconstructed $S_\text{geo}$ plotted against the true $E_\text{em}^\text{MC}$. The right columns of each panel display the deviation of reconstructed $E_\text{em}$ from the true value with corresponding resolution and bias plotted against $\text{log}_{10}(E_\text{em}^\text{MC} / \mathrm{eV})$. \textbf{Left:} GRAND@Auger. \textbf{Right:} GP300.}
    \label{fig:recon}
\end{figure}

Fig.~~\ref{fig:recon} shows the reconstruction performance of $S_\text{geo}$ and $E_\text{em}$ intrinsic to our approach (i.e., for an idealized situation where star-shape simulations are available and no experimental uncertainties are imprinted).
We show the reconstruction with all LDF shape parameters left free.
For both the GRAND@Auger and GP300 sites, we achieve a resolution of $<5\%$ with no significant biases.

\section{Conclusions}
\label{sec:conclusions}

We have adapted the signal model from reference \cite{Model-paper} to the $50-200\,\mathrm{MHz}$ frequency band for both the GRAND@Auger site in Argentina and the GP300 site in China. We have refitted the parameterisation of the charge excess fraction. In addition, we have modified the functional form of the LDF with a new parameter for the inside of the Cherenkov radius, and changed the way the parameter $r_0$ is determined.
For the GRAND@Auger site, these changes have been sufficient to produce an accurate signal model and functional energy reconstruction with a resolution $<5\%$ with no significant biases.
For the GP300 site in China, the stronger magnetic field leads to coherence loss in the radio emission.
To address this, we have adapted the density correction term used to calculate the corrected radiation energy $S_\text{geo}$.
As a result, the energy reconstruction for this site achieves a resolution of $<5\%$ as well.
For the next step, we will make last optimisations to the model and parameterise the LDF parameters with $d_\text{max}$. Finally, we will benchmark the reconstruction on realistic simulations with a sparse antenna grid and added noise.

\section*{Acknowledgments}
This work is part of the NUTRIG project, supported by the Agence Nationale de la Recherche (ANR-21-CE31-0025; France) and the Deutsche Forschungsgemeinschaft (DFG; Projektnummer 490843803; Germany).
Computations were performed using the resources of the HoreKa supercomputer, funded by the Ministry of Science, Research and the Arts Baden-W{\"u}rttemberg and by the Federal Ministry of Education and Research.

\bibliographystyle{JHEP}
\bibliography{bibliography}

\end{document}